\documentclass[%
reprint,
superscriptaddress,
showpacs,
amsmath,amssymb,
aps,
prl,
]{revtex4-2}

\usepackage{amsmath}
\usepackage{graphicx}
\usepackage{braket}
\usepackage{amssymb}
\usepackage{float}
\usepackage{color}
\usepackage[colorlinks=true,linkcolor=blue,anchorcolor=blue,citecolor=blue,urlcolor=blue]{hyperref}
\usepackage{bm}
\usepackage{comment}
\begin{document}
\preprint{APS/123-QED}

\title{Attosecond Access to the Quantum Noise of Light}
\author{En-Rui Zhou}
\thanks{These authors contributed equally to this work.}
\affiliation{State Key Laboratory of Dark Matter Physics, Key Laboratory for Laser Plasmas (Ministry of Education) and School of Physics and Astronomy, Collaborative Innovation Center of IFSA (CICIFSA), Shanghai Jiao Tong University, Shanghai 200240, China}
\author{Yi-Jia Mao}
\thanks{These authors contributed equally to this work.}
\affiliation{State Key Laboratory of Dark Matter Physics, Key Laboratory for Laser Plasmas (Ministry of Education) and School of Physics and Astronomy, Collaborative Innovation Center of IFSA (CICIFSA), Shanghai Jiao Tong University, Shanghai 200240, China}
\affiliation{Tsung-Dao Lee Institute, Shanghai Jiao Tong University, Shanghai 201210, China}
\author{Pei-Lun He}
\email{peilunhe@sjtu.edu.cn}
\affiliation{State Key Laboratory of Dark Matter Physics, Key Laboratory for Laser Plasmas (Ministry of Education) and School of Physics and Astronomy, Collaborative Innovation Center of IFSA (CICIFSA), Shanghai Jiao Tong University, Shanghai 200240, China}
\author{Feng He}
\affiliation{State Key Laboratory of Dark Matter Physics, Key Laboratory for Laser Plasmas (Ministry of Education) and School of Physics and Astronomy, Collaborative Innovation Center of IFSA (CICIFSA), Shanghai Jiao Tong University, Shanghai 200240, China}
\affiliation{Tsung-Dao Lee Institute, Shanghai Jiao Tong University, Shanghai 201210, China}

\date{\today}

\begin{abstract}
Bright squeezed light has entered strong-field atomic physics, creating an immediate need to characterize the quantum field delivered to the target on subcycle timescales. Here we show that attosecond streaking maps the coherent displacement and phase-sensitive covariance of a displaced squeezed field onto distinct harmonics: the mean photoelectron momentum follows an $\omega$-periodic shift of the spectral center, whereas the momentum variance exhibits a $2\omega$-periodic breathing. A Feynman--Vernon formulation represents Gaussian quantum light by stochastic vector-potential trajectories, enabling Coulomb-resolved TDSE simulations with a finite XUV gate and without explicit photon-state propagation.
A coherent-state reference calibrates the streaking phase and variance background, enabling retrieval of the coherent amplitude and phase together with the squeezed-noise amplitude and squeezing phase. This establishes attosecond streaking as an in situ, gas-phase diagnostic of bright squeezed light.
\end{abstract}

\maketitle

Advances in bright-squeezed-vacuum (BSV) generation and photon statistics have made intense quantum light experimentally accessible~\cite{exp_bsv_1,exp_bsv_2,exp_bsv_3,lyu2026generation}. 
This progress has enabled the first strong-field atomic tunnelling experiment with BSV~\cite{jiang2026nonlinear} and stimulated studies of quantum-light-driven ionization~\cite{qati1,qati2,qati3,qati4,liu2026Strong-field}, continuum-electron dynamics~\cite{even2024motion,heimerl2025quantum}, high-harmonic generation~\cite{hhg_bsv_solid,theidel2024evidence,lemieux2025photon,oht3,tzur2025attosecond}, and molecular dissociation~\cite{dis}. Yet ultrafast light--matter dynamics is governed by the time-dependent quantum field at the target rather than by source-level statistics alone~\cite{review1,review2,jitter1}. Residual seeding, propagation, and imperfect mode matching can introduce a coherent displacement or modify the effective mode covariance of a nominal BSV~\cite{loudon1987squeezed,breitenbach1997measurement,vahlbruch2016squeezed}; determining its local optical phase, squeezing angle, and residual displacement therefore requires a phase-resolved in situ diagnostic.

Photon counting and intensity correlations, such as $g^{(2)}$, characterize photon-number statistics but not the optical phase or subcycle quadrature waveform~\cite{hbt1956,glauber1963}. Homodyne methods resolve field quadratures but are not naturally compatible with attosecond gating under strong-field conditions~\cite{oht1,oht2}. 
Electro-optic and all-optical sampling achieve subcycle resolution but rely on calibrated material responses and operate over limited intensity ranges~\cite{tr1,tr2,tr3,sennary2025attosecond,oht4}. These limitations motivate an in situ, gas-phase probe in which the atom directly samples the driving quantum field.

Attosecond streaking spectroscopy provides such a route. An isolated extreme-ultraviolet (XUV) pulse ionizes an atom in the presence of a dressing infrared (IR) field, and the emitted photoelectron acquires, to leading order, a momentum shift determined by the vector potential at the ionization time~\cite{streak_1,streak_2,streak_3,streak_5,streak_6,streak_4,PhysRevLett.88.173904}. In conventional streaking, the delay-dependent first moment of the photoelectron spectrum reconstructs the coherent vector potential, whereas the spectral width is usually treated as a finite-XUV or instrumental background. Here we show that, for a quantum dressing field, the first moment and second central moment of the same delay-resolved spectrum separate the coherent and fluctuation sectors: the mean momentum carries the $\omega$-periodic coherent response, whereas the variance carries the phase-sensitive $2\omega$ response of the squeezed covariance. We derive this mapping using a Feynman--Vernon influence functional and validate it with TDSE simulations that retain the Coulomb potential and a finite XUV gate. This reduced-dynamics construction complements fully quantized simulations and benchmarks of quantum-light-driven electron and high-harmonic dynamics~\cite{full_quantum_hhg,mao2025benchmarking,liu2026theory}, while avoiding explicit propagation of the coupled electron--photon state.
An otherwise identical coherent-state trace calibrates the atomic streaking phase and variance background, enabling quantitative reconstruction of the amplitudes and phases of both field components.

We consider the electron dynamics governed by the minimal-coupling Hamiltonian
\begin{equation}
\hat H(t)
=
\frac{\bigl[\hat{\bm p}+\hat{\bm A}(t)\bigr]^2}{2}
+
V_{\mathrm C}(\hat{\bm x}),
\label{eq:fullH}
\end{equation}
where $\hat{\bm A}(t)$ is the quantized vector potential in the dipole approximation and $V_{\mathrm C}(\hat{\bm x})$ the Coulomb potential. Atomic units are used throughout unless stated otherwise. Integrating out the photonic degrees of freedom yields the Feynman--Vernon influence functional~\cite{Feynman,parikh2021quantum}, which fully determines the photon-traced electron dynamics. For Gaussian quantum light, such as squeezed coherent and thermal states, the influence functional separates into a coherent driving term, a fluctuation kernel, and a dissipative radiation-reaction kernel. In the regime considered here, the dissipative contribution is negligible, while the fluctuation kernel can be represented through a Hubbard--Stratonovich transformation as a Gaussian ensemble of stochastic fields~\cite{1957SPhD....2..416S,PhysRevLett.3.77,Parikh:2021nqj,paraguassu2024quantum}. The reduced electron dynamics can therefore be written as an ensemble of evolutions generated by
\begin{equation}
\hat{H}_{\bm{\mathcal N}}(t)
=
\frac{
\bigl[
\hat{\bm p}
+
\bm A_{\mathrm{cl}}(t)
+
\bm{\mathcal N}(t)
\bigr]^2
}{2}
+
V_{\mathrm C}(\hat{\bm x}).
\label{eq:Hstoch}
\end{equation}
Here $\bm A_{\mathrm{cl}}(t)=\langle\hat{\bm A}(t)\rangle$ is the coherent mean field, and $\bm{\mathcal N}(t)$ is a zero-mean Gaussian stochastic field satisfying
\begin{equation}
\langle \mathcal N_i(t)\rangle=0,
\qquad
\langle \mathcal N_i(t)\mathcal N_j(t')\rangle
=
\nu_{ij}(t,t'),
\label{eq:noisecorr}
\end{equation}
where $\nu_{ij}(t,t')$ is the symmetrically ordered covariance of the vector-potential fluctuations~\cite{supp}, with $i,j=x,y,z$. 
For Gaussian states with positive Wigner functions and negligible radiation reaction, the reduced dynamics therefore admits an exact positive trajectory representation weighted by the Wigner distribution. This weighting is fixed by the symmetric ordering of the influence functional, whereas direct Husimi-$Q$ sampling requires the corresponding vacuum-fluctuation correction~\cite{qrep_hhg1,qrep_hhg2}. Comparison with fully quantized calculations confirms that the Wigner prescription reproduces the photon-traced dynamics, while uncorrected Husimi-$Q$ sampling introduces an additional vacuum contribution~\cite{supp}. Further details are provided in the End Matter and Supplemental Material.

\begin{figure}
\centering
\includegraphics[width=0.48\textwidth]{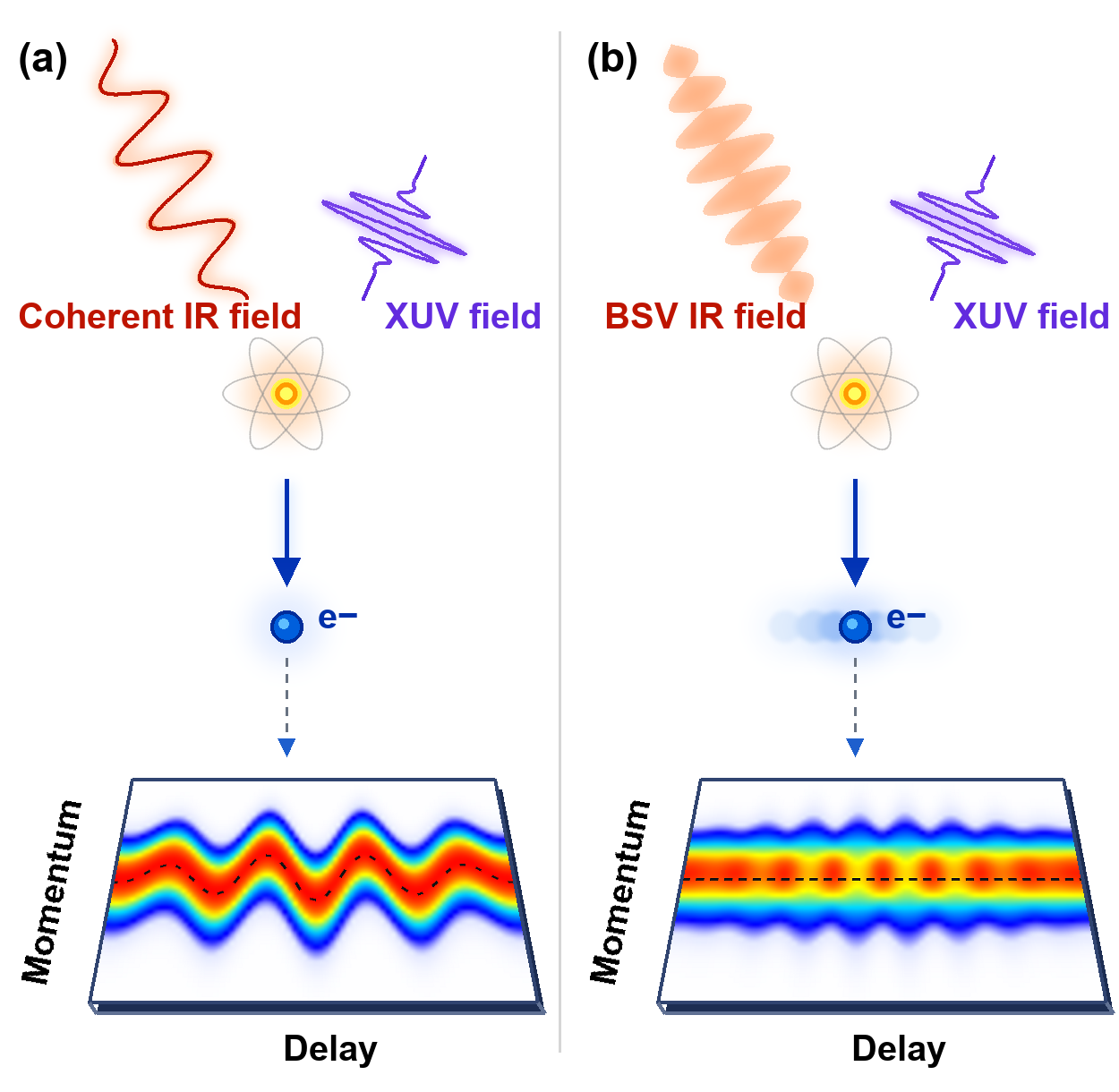}
\caption{
Principle of attosecond streaking driven by quantum light. An isolated XUV pulse ionizes an atom and releases a photoelectron with central momentum $p_0=\sqrt{2(\omega_{\mathrm{XUV}}-I_p)}$, whose final momentum samples the effective vector potential $A_{\mathrm{cl}}(\tau)+\mathcal N(\tau)$ at the ionization delay $\tau$. (a)~For a coherent-state IR field ($r=0$), the center of the photoelectron spectrum oscillates at the carrier frequency $\omega$, while the variance follows only the slowly varying pulse envelope. (b)~For bright squeezed vacuum, corresponding to the $\alpha=0$ limit of a squeezed coherent state, the mean streaking shift vanishes, while the spectral width exhibits a phase-locked $2\omega$ breathing determined by the squeezing phase $\theta$.
}
\label{fig:streaking_schematic}
\end{figure}

We now specialize the stochastic formulation to the attosecond-streaking configuration illustrated in Fig.~\ref{fig:streaking_schematic}. Projecting Eq.~\eqref{eq:Hstoch} onto the common polarization axis of the IR and XUV fields reduces the streaking dynamics to the longitudinal component. For the numerical examples, we consider a narrowband pulse dominated by a single collective wave-packet mode, formed from a superposition of nearby plane-wave modes and described by the slowly varying envelope $f(t)$~\cite{qV}. This effective mode is prepared in a squeezed coherent state,
\(
|\alpha,\xi\rangle=\hat D(\alpha)\hat S(\xi)|0\rangle,
\)
with $\alpha=|\alpha|e^{i\phi}$ and $\xi=re^{i\theta}$. Its coherent vector potential is
\begin{equation}
A_{\mathrm{cl}}(t)
=
\frac{2\mathcal{E}_V}{\omega}\,
f(t)\,|\alpha|\cos(\omega t-\phi),
\label{eq:Acl}
\end{equation}
and its fluctuation kernel is
\begin{align}
\nu(t,t')
&=
\frac{\mathcal{E}_V^2}{\omega^2}f(t)f(t')
\Bigl\{
\cosh(2r)\cos[\omega(t-t')] \notag\\
&\quad
-\sinh(2r)\cos[\omega(t+t')-\theta]
\Bigr\}.
\label{eq:kernel}
\end{align}
Here $\mathcal{E}_V=\sqrt{2\pi\omega/V}$ is the single-photon field amplitude associated with the normalized effective mode, and $V$ is the corresponding quantization volume. The first term in Eq.~\eqref{eq:kernel} contains the vacuum contribution and the phase-insensitive part of the squeezed fluctuations, whereas the second is the phase-sensitive, $t+t'$-dependent anomalous correlation. At equal times,
\begin{equation}
\nu(\tau,\tau)
=
\frac{\mathcal E_V^2}{\omega^2}
f^2(\tau)
\left[
\cosh(2r)
-
\sinh(2r)\cos(2\omega\tau-\theta)
\right],
\label{eq:nu_equal}
\end{equation}
so the squeezing strength $r$ determines the modulation amplitude, while the squeezing phase $\theta$ determines its delay phase.
The phase-locked $2\omega$ component of Eq.~\eqref{eq:nu_equal} therefore isolates the phase-sensitive squeezed covariance.

A classical XUV pulse with central frequency $\omega_{\mathrm{XUV}}$ ionizes an atom with ionization potential $I_p$ in the presence of the quantum IR dressing field. Within the sudden-ionization approximation~\cite{PhysRevLett.88.173904,streak_4}, the electron is released at the pump--probe delay $\tau$. In the absence of the IR field, its central continuum momentum is
\(
p_0=\sqrt{2(\omega_{\mathrm{XUV}}-I_p)}
\).
We account for the finite duration and bandwidth of the ionizing pulse through an initial momentum fluctuation $\delta p_{\mathrm{XUV}}$, with $\langle\delta p_{\mathrm{XUV}}\rangle=0$ and $\langle\delta p_{\mathrm{XUV}}^2\rangle=\sigma_0^2$, where $\sigma_0^2$ is the delay-independent variance. Neglecting the post-ionization Coulomb interaction, the stochastic Hamiltonian in Eq.~\eqref{eq:Hstoch} yields
\begin{equation}
p(\tau)
=
p_0
+
\delta p_{\mathrm{XUV}}
-
A_{\mathrm{cl}}(\tau)
-
\mathcal N(\tau).
\label{eq:streakingprinciple}
\end{equation}
Equation~\eqref{eq:streakingprinciple} provides the analytical Coulomb-free mapping, whereas the full TDSE calculations below retain both the Coulomb potential and the finite XUV gate.

Averaging Eq.~\eqref{eq:streakingprinciple} over the initial XUV-induced momentum distribution and the stochastic field gives
\begin{equation}
\langle p(\tau)\rangle
=
p_0-A_{\mathrm{cl}}(\tau),
\label{eq:mean}
\end{equation}
and
\begin{equation}
\langle\Delta p^2(\tau)\rangle
=
\sigma_0^2+\nu(\tau,\tau).
\label{eq:var}
\end{equation}
The delay-dependent mean momentum therefore measures the coherent mean field, whereas the spectral variance measures the equal-time field-fluctuation covariance. Although the numerical examples below use a single dominant collective mode, Eqs.~\eqref{eq:mean} and~\eqref{eq:var} remain valid for multimode Gaussian fields, with $A_{\mathrm{cl}}(\tau)$ and $\nu(\tau,\tau)$ interpreted as the projected mean and equal-time covariance~\cite{supp}.

Substituting Eqs.~\eqref{eq:Acl} and~\eqref{eq:nu_equal} into Eqs.~\eqref{eq:mean} and~\eqref{eq:var} makes the two limiting signatures in Fig.~\ref{fig:streaking_schematic} transparent. For a coherent state [Fig.~\ref{fig:streaking_schematic}(a)], $r=0$, so the mean momentum follows $\cos(\omega\tau-\phi)$, while the IR-induced variance reduces to the envelope-following vacuum term
\(
\nu_{\mathrm{coh}}(\tau,\tau)
=
\frac{\mathcal E_V^2}{\omega^2}f^2(\tau).
\)
In the bright-squeezed-vacuum limit [Fig.~\ref{fig:streaking_schematic}(b)], $\alpha=0$, so the mean streaking shift vanishes and the signal appears as a $\cos(2\omega\tau-\theta)$ modulation of the spectral width. This phase-locked modulation originates from the anomalous, $t+t'$-dependent covariance. By contrast, coherent and thermal states, as well as Fock states at the level of their two-time covariance, contribute only phase-insensitive, envelope-following broadening at equal times. 

\begin{figure}
    \centering
    \includegraphics[width=0.5\textwidth]{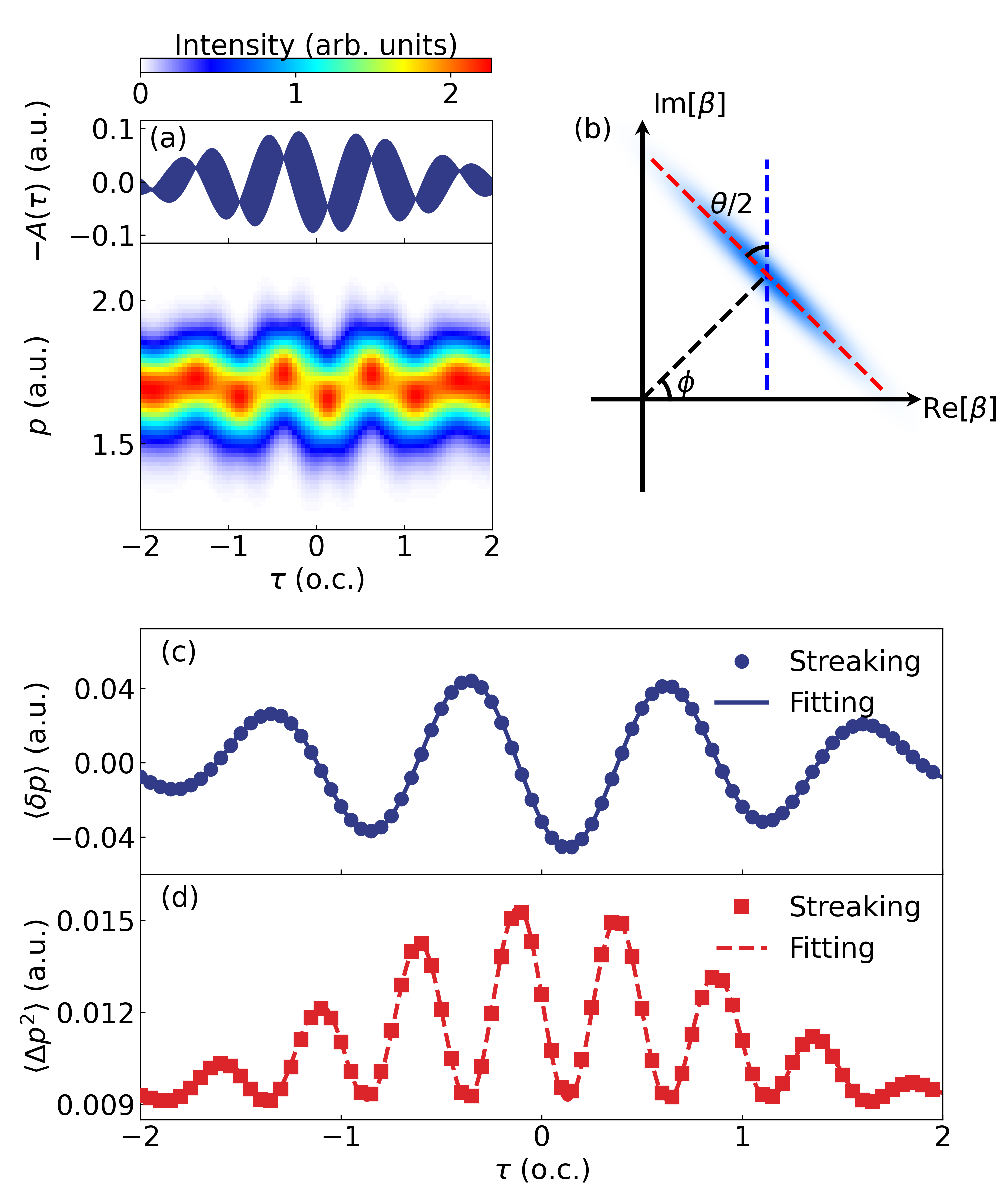}
\caption{Characterization of a squeezed coherent IR field by attosecond streaking. (a)~Time-domain waveform of the squeezed coherent IR field (upper panel) and the corresponding attosecond streaking trace (lower panel). The total effective intensity is $I_c+I_s=1\times10^{12}\,\mathrm{W/cm^2}$, with coherent-to-squeezed intensity ratio $I_c/I_s=1/3$, coherent phase $\phi=\pi/4$, and squeezing phase $\theta=\pi/2$. (b)~Phase-space representation of the displaced squeezed state, showing the coherent displacement and the orientation of the squeezed quadrature. (c)~Mean momentum shift $\langle\delta p(\tau)\rangle$, carrying the $\omega$-periodic coherent signature, and (d)~momentum variance $\langle\Delta p^2(\tau)\rangle$, carrying the $2\omega$-periodic squeezing signature. The blue circles in (c) and red squares in (d) are extracted from the streaking trace in (a), while the corresponding solid blue and dashed red curves are fits to Eqs.~\eqref{pfit} and~\eqref{p2fit}. After calibration with the coherent-state reference, the fits yield $\phi=0.786\,\mathrm{rad}$ and $\theta=1.572\,\mathrm{rad}$, compared with the input values $\pi/4\approx0.785\,\mathrm{rad}$ and $\pi/2\approx1.571\,\mathrm{rad}$, respectively.
}
\label{fig:squeezed_streaking}
\end{figure}

Having established these limiting signatures, we now formulate a calibrated retrieval protocol for a general squeezed coherent IR field with both $\alpha\neq0$ and $r\neq0$. We define the coherent amplitude $E_c=2\mathcal E_V|\alpha|$ and, in the bright limit $r\gg1$, the amplified squeezed-noise amplitude $E_s\simeq\mathcal E_Ve^{r}$, with corresponding effective intensities $I_c\propto E_c^2$ and $I_s\propto E_s^2$. 
For the pulse parameters considered here, deviations from the instantaneous Volkov mapping due to Coulomb scattering~\cite{eisenbud,wigner,smith,delay} and finite-gate averaging are accurately captured by a common effective time offset $\Delta t$, calibrated using an otherwise identical coherent-state reference.
Defining
\(
\delta=\omega\Delta t
\)
and
\(
\langle\delta p(\tau)\rangle
=
\langle p(\tau)\rangle-p_0,
\)
the calibrated fitting formulas are
\begin{equation}
\langle\delta p(\tau)\rangle
=
-\frac{E_c}{\omega}
f(\tau)
\cos\!\left(\omega\tau-\phi-\delta\right),
\label{pfit}
\end{equation}
and
\begin{equation}
\langle\Delta p^2(\tau)\rangle
=
\frac{E_s^2}{2\omega^2}
f^2(\tau)
\left[
1-\cos\!\left(2\omega\tau-\theta-2\delta\right)
\right]
+\sigma_0^2.
\label{p2fit}
\end{equation}
The coherent-state reference fixes $\delta$ from the phase of its mean-momentum trace and $\sigma_0^2$ from its variance after subtracting the envelope-following coherent-state contribution. With these calibration parameters fixed, Eqs.~\eqref{pfit} and~\eqref{p2fit} retrieve $E_c$, $E_s$, $\phi$, and $\theta$ from the squeezed coherent trace. The directly fitted phases consequently satisfy
\(
\phi_{\mathrm{streak}}=\phi+\delta
\)
and
\(
\theta_{\mathrm{streak}}=\theta+2\delta
\).

We implement this protocol in TDSE simulations using an isolated \(54\,\mathrm{eV}\) XUV pulse to ionize the atom in the presence of an \(800\,\mathrm{nm}\) quantum IR dressing field. As a representative squeezed coherent case, we fix the total effective intensity to \(I_c+I_s=1\times10^{12}\,\mathrm{W/cm^2}\) and choose \(I_c/I_s=1/3\), with \(\phi=\pi/4\) and \(\theta=\pi/2\). Figure~\ref{fig:squeezed_streaking}(a) shows the time-domain waveform of the squeezed coherent IR field in the upper panel and the corresponding attosecond streaking trace in the lower panel. The phase-space representation in Fig.~\ref{fig:squeezed_streaking}(b) displays the coherent displacement and the orientation of the squeezed quadrature. 
The spectral center of the streaking trace oscillates with delay and follows the negative coherent vector potential, while its spectral width exhibits the phase-sensitive \(2\omega\) breathing predicted by Eq.~\eqref{p2fit}. The mean momentum shift and momentum variance extracted from this trace are shown in Figs.~\ref{fig:squeezed_streaking}(c) and~\ref{fig:squeezed_streaking}(d), respectively, together with fits to Eqs.~\eqref{pfit} and~\eqref{p2fit}.

For the coherent-state reference, we fit the mean streaking trace
and obtain the apparent streaking phase
\(\phi_{\mathrm{streak}}^{(\mathrm{ref})}\).
The phase offset is then determined by comparing it with the known input phase,
$
\delta
=
\phi_{\mathrm{streak}}^{(\mathrm{ref})}
-
\phi_{\mathrm{in}}^{(\mathrm{ref})}
=
0.032\,\mathrm{rad}.
$ Using this calibrated offset, the squeezed coherent trace yields $\phi_{\mathrm{streak}}=0.818\,\mathrm{rad}$ and $\theta_{\mathrm{streak}}=1.636\,\mathrm{rad}$.
Subtracting the coherent-reference phase correction gives
$\phi=\phi_{\mathrm{streak}}-\delta=0.786\,\mathrm{rad}$ and $\theta=\theta_{\mathrm{streak}}-2\delta=1.572\,\mathrm{rad}$,
in good agreement with the input field parameters. 
The TDSE-extracted variance phase further confirms that the $2\omega$ channel acquires the phase offset $2\delta$ within numerical accuracy.
The fitted amplitudes determine $E_c$ and $E_s$, and hence the relative weights of the coherent displacement and squeezed fluctuations. The independently calibrated background $\sigma_0^2$ accounts for the delay-independent part of the spectral width, confirming that this contribution is set primarily by the finite XUV bandwidth rather than by the squeezed IR fluctuations~\cite{supp}.

\begin{figure}
    \centering
    \includegraphics[width=0.48\textwidth]{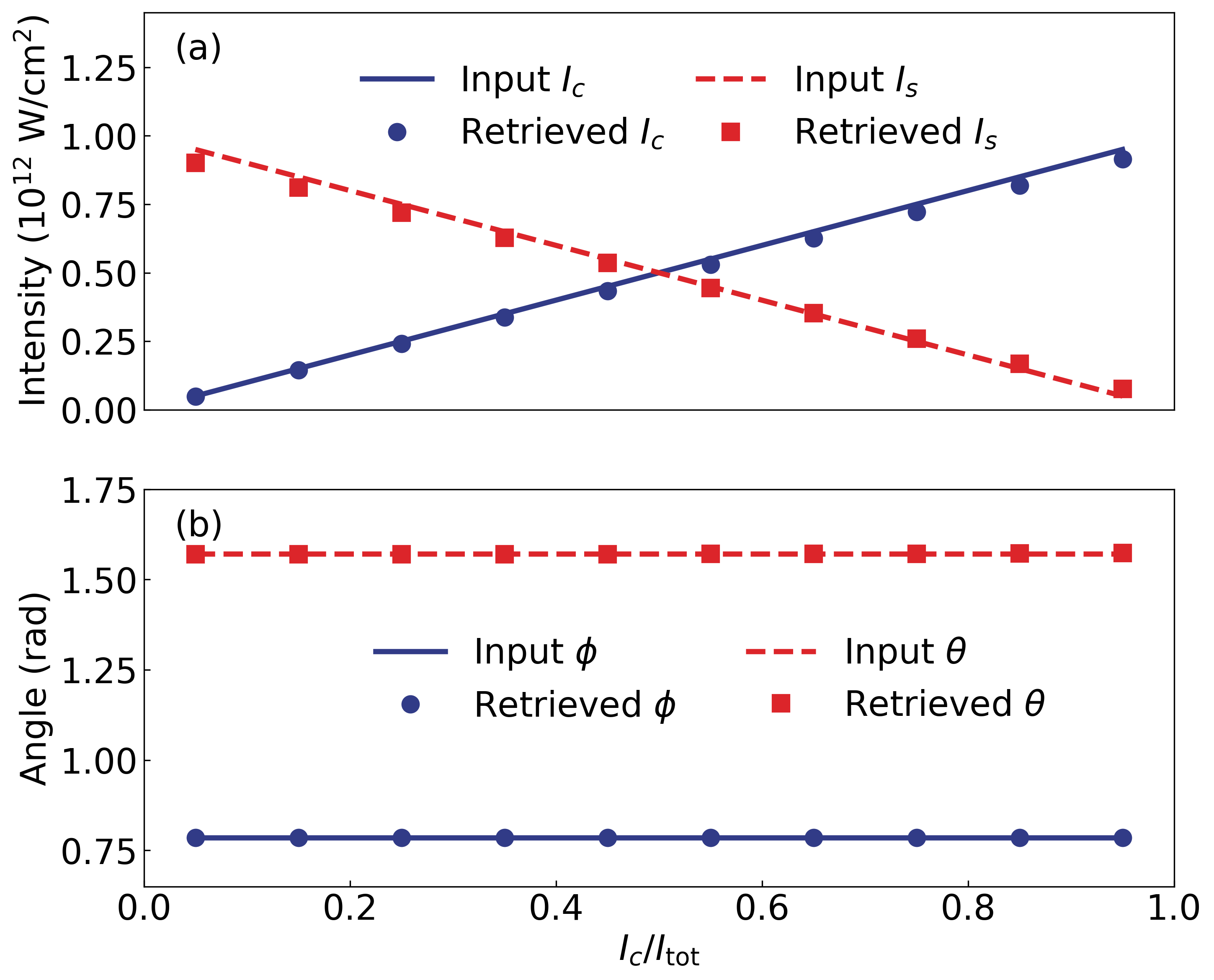}
\caption{Retrieval of the squeezed-coherent-field parameters versus the coherent intensity fraction $I_c/I_{\mathrm{tot}}$, for fixed $I_{\mathrm{tot}}=I_c+I_s=1\times10^{12}\,\mathrm{W/cm^2}$. (a)~Input coherent and squeezed intensities, $I_c$ (blue solid line) and $I_s$ (red dashed line), and their retrieved values (circles and squares). (b)~Input phases $\phi=\pi/4$ and $\theta=\pi/2$ and the corresponding retrieved values. The retrieval reproduces all four input parameters across the scanned range. Other parameters are the same as in Fig.~\ref{fig:squeezed_streaking}.}
\label{fig:para_scan}
\end{figure}

Having demonstrated quantitative retrieval for the representative state in Fig.~\ref{fig:squeezed_streaking}, we next test the robustness of the protocol against variations in the field composition and experimental conditions. At fixed $I_{\mathrm{tot}}=I_c+I_s=1\times10^{12}\,\mathrm{W/cm^2}$, Fig.~\ref{fig:para_scan}(a) compares the input intensities $I_c$ and $I_s$ with their retrieved values as the coherent fraction $I_c/I_{\mathrm{tot}}$ is varied, while Fig.~\ref{fig:para_scan}(b) shows the corresponding comparison for the fixed phases $\phi=\pi/4$ and $\theta=\pi/2$. The retrieval reproduces all four parameters throughout the scanned range, demonstrating that the protocol remains quantitatively stable as the relative weights of the coherent and squeezed components change. 
Lowering the total IR intensity primarily reduces the modulation contrast while preserving the harmonic phase relations on which the retrieval relies~\cite{supp}. Combined with the sub-eV energy resolution already demonstrated in attosecond photoelectron spectroscopy~\cite{luo2023ultra}, our simulations give a resolution-based sensitivity estimate of order $10^{8}\,\mathrm{W/cm^2}$~\cite{supp}. 
Focal-volume averaging and a carrier-envelope-phase jitter of $50\,\mathrm{mrad}$ primarily reduce the modulation contrast, while changing the retrieved phases by less than $10^{-3}\,\mathrm{rad}$ within the simulated conditions~\cite{supp}. The relevant experimental criterion is therefore that the $\omega$ and $2\omega$ moment components remain resolvable within the available spectral resolution and counting statistics.

These results establish the protocol as an in situ diagnostic of the coherent and squeezed components of the quantum field delivered to the interaction region. Applied to recent BSV-driven strong-field experiments~\cite{jiang2026nonlinear}, it can test whether the field at the target remains an undisplaced squeezed vacuum or carries a residual coherent displacement, without assuming an ideal BSV at the outset. The retrieval further determines the displacement amplitude and phase together with the phase-sensitive squeezed covariance. Additional stationary broadening can signal deviations from a pure displaced-squeezed model. Distinguishing possible photon-state contributions, such as thermal and large-\(n\) Fock admixtures, requires higher spectral moments, as discussed in the End Matter.

In conclusion, attosecond streaking provides a phase-sensitive, subcycle probe of bright squeezed light. The first two moments of a single delay-resolved streaking trace separately retrieve the coherent displacement and the phase-sensitive squeezed covariance. With an otherwise identical coherent-state trace as a reference, both components can be reconstructed quantitatively in the presence of the Coulomb potential and a finite XUV gate, establishing streaking as an in situ, gas-phase complement to optical homodyne detection. 
More broadly, for Gaussian radiation states with positive Wigner functions and negligible radiation reaction, the Feynman--Vernon formulation maps the photon-traced dynamics onto a Wigner-weighted ensemble of classical-field-driven TDSE evolutions, avoiding explicit propagation of the photonic degrees of freedom.  For multimode fields, Fourier analysis of the delay-dependent mean and variance may provide mode-resolved access to the projected mean field and covariance when the contributing modes are spectrally distinguishable. Extensions retaining the dissipative influence kernel and photon backaction may further open a route to effective non-Hermitian strong-field dynamics.


\textit{Note added.---} During the revision of this manuscript, we became aware of an independent related work on attosecond metrology of bright quantum light~\cite{stammer2026attosecond}. While that study emphasizes squeezing certification, the present work addresses the calibrated reconstruction of a general displaced squeezed field under Coulomb-resolved strong-field conditions.

\textit{Acknowledgments}\textemdash
This work was supported by National Natural Science Foundation of China (NSFC) (Grant Nos.\ 12450405, 12274294, 12574378, 12574377). The computations in this paper were run on the Siyuan-1 cluster supported by the Center for High Performance Computing at Shanghai Jiao Tong University. P.-L.~H.\ acknowledges support from the Pujiang Program of the Shanghai Baiyulan Talent Plan (Grant No.~24PJA046), the Xiaomi Young Scholar Program, the Shanghai Jiao Tong University 2030 Initiative, and the Yangyang Development Fund. Y.-J.~M.\ is supported by T.D.\ Lee Scholarship.


\bibliography{ref}

\clearpage
\begin{widetext}
\section*{End Matter}

Here we summarize the influence-functional formulation underlying the stochastic quantum-light sampling used in the main text. Detailed derivations are given in the Supplemental Material~\cite{supp}.

\textit{Influence functional and phase-space representations.---}
We first consider a single monochromatic radiation mode as a building block of
the effective wave-packet description used in the main text. In the
closed-time-path formulation, tracing out this photonic mode gives the
Feynman--Vernon influence functional \(F_\gamma[\Delta\dot{x}]\), where
\(\Delta\dot{x}(t)\equiv
[\dot{\bm{x}}(t)-\dot{\bm{x}}'(t)]\cdot\bm\varepsilon\). Neglecting
radiation reaction and adopting the dipole approximation,
\begin{equation}
F_\gamma[\Delta\dot{x}]
=
\mathrm{Tr}\!\left[
\rho_\gamma
\hat{D}\!\left(\gamma[\Delta\dot{x}]\right)
\right],
\qquad
\gamma[\Delta\dot{x}]
=
-\frac{i\mathcal{E}_V}{\omega}
\int_{t_i}^{t_f}\!\mathrm{d}t\,
\Delta\dot{x}(t)e^{i\omega t},
\label{em:F_general}
\end{equation}
with
\(\hat{D}(\gamma)=\exp(\gamma\hat{a}^{\dagger}-\gamma^*\hat{a})\).
This is the symmetrically ordered characteristic function of the photon state,
whose phase-space Fourier partner is the Wigner function~\cite{quantum_optics}. Therefore
\begin{equation}
F_\gamma[\Delta\dot{x}]
=
\int\!\mathrm{d}^2\beta\,
W_\rho^{(\mathrm W)}(\beta)
\exp\!\left(
\gamma\beta^*
-
\gamma^*\beta
\right),
\label{em:Wigner_Fourier}
\end{equation}
where \(W_\rho^{(\mathrm W)}(\beta)\) is the Wigner function. When
\(W_\rho^{(\mathrm W)}(\beta)\ge0\), Eq.~\eqref{em:Wigner_Fourier} gives an
exact positive fluctuation weight.

This also fixes the ordering of the noise kernel. For positive Wigner weights,
the field average gives
\begin{equation}
\nu(t,t')
=
\int\!d^2\beta\,
W_\rho^{(\mathrm W)}(\beta)\,
\delta A(\beta,t)\delta A(\beta,t')
=
\frac{1}{2}
\left\langle
\delta\hat A(t)\delta\hat A(t')
+
\delta\hat A(t')\delta\hat A(t)
\right\rangle ,
\label{em:nu_sym}
\end{equation}
where \(\delta A(\beta,t)=A(\beta,t)-A_{\mathrm{cl}}(t)\). Thus the exact
kernel derived from the influence functional is a symmetrically ordered
two-time field correlator.

For a squeezed coherent state
\(|\alpha,\xi\rangle=\hat D(\alpha)\hat S(\xi)|0\rangle\), with
\(\alpha=|\alpha|e^{i\phi}\) and \(\xi=re^{i\theta}\), the positive Wigner
weight is
\begin{equation}
W_{\alpha,\xi}^{(\mathrm W)}(\beta)
=
\frac{2}{\pi}
\exp\!\left[
-2e^{-2r}u_\beta^2
-
2e^{2r}v_\beta^2
\right],
\label{em:W_sq}
\end{equation}
where
\begin{align}
u_\beta
&=
(\mathrm{Re}\beta-|\alpha|\cos\phi)\sin\frac{\theta}{2}
-
(\mathrm{Im}\beta-|\alpha|\sin\phi)\cos\frac{\theta}{2},
\notag\\
v_\beta
&=
(\mathrm{Re}\beta-|\alpha|\cos\phi)\cos\frac{\theta}{2}
+
(\mathrm{Im}\beta-|\alpha|\sin\phi)\sin\frac{\theta}{2}.
\end{align}
It gives
\[
A_{\mathrm{cl}}(t)
=
\frac{\mathcal{E}_V}{\omega}
\left(
\alpha e^{-i\omega t}
+
\alpha^*e^{i\omega t}
\right)
\]
and
\begin{equation}
\nu_{\xi}(t,t')
=
\frac{\mathcal{E}_V^2}{\omega^2}
\left[
\cosh(2r)\cos\omega(t-t')
-
\sinh(2r)\cos\!\left(\omega(t+t')-\theta\right)
\right].
\label{em:nu_sq}
\end{equation}
The \(t+t'\)-dependent anomalous term carries the squeezing phase and produces
the \(2\omega\) modulation of the streaking variance.

For a thermal state with mean photon number \(\bar n\),
\begin{equation}
W_{\mathrm{th}}^{(\mathrm W)}(\beta)
=
\frac{1}{\pi(\bar n+\frac{1}{2})}
\exp\!\left[
-\frac{|\beta|^2}{\bar n+\frac{1}{2}}
\right],
\label{em:W_th}
\end{equation}
and
\begin{equation}
\nu_{\mathrm{th}}(t,t')
=
\frac{\mathcal{E}_V^2}{\omega^2}
(2\bar n+1)\cos\left[\omega(t-t')\right].
\label{em:nu_th}
\end{equation}
This \(t-t'\)-dependent kernel gives envelope-following broadening but no
phase-sensitive \(2\omega\) modulation.

If the Wigner function is not positive, Eq.~\eqref{em:Wigner_Fourier} remains
exact but is not a probability average. One can instead rewrite the influence
functional in antinormal order,
\begin{align}
F_\gamma[\Delta\dot{x}]
=
e^{|\gamma|^2/2}
\mathrm{Tr}\!\left[
\rho_\gamma
e^{-\gamma^*\hat a}
e^{\gamma\hat a^\dagger}
\right].
\label{em:F_antinormal}
\end{align}
The trace is the antinormally ordered characteristic
function, i.e. the phase-space Fourier transform of the Husimi \(Q\) function~\cite{quantum_optics}.
Therefore
\begin{equation}
F_\gamma[\Delta\dot{x}]
=
e^{|\gamma|^2/2}
\int\!\mathrm{d}^2\beta\,
Q_\rho(\beta)
\exp\!\left(
\gamma\beta^*
-
\gamma^*\beta
\right),
\label{em:Q_Fourier}
\end{equation}
where
\(
Q_\rho(\beta)
=
\frac{\langle\beta|\rho_\gamma|\beta\rangle}{\pi}\)
is the Husimi \(Q\) function.  Although \(Q_\rho\) is positive for any photon
state, direct \(Q\)-sampling without the factor \(e^{|\gamma|^2/2}\) neglects
a vacuum-fluctuation-scale correction, and the corresponding two-time moments
are antinormally ordered.

For a Fock state, the Wigner function is sign-changing for \(n\ge1\), so the
exact Wigner representation cannot be used as a positive sampling weight.
Within the positive phase-space sampling scheme, one therefore uses the
antinormal formula above and samples the Husimi distribution
\begin{equation}
Q_n(\beta)
=
\frac{\langle\beta|n\rangle\langle n|\beta\rangle}{\pi}
=
\frac{e^{-|\beta|^2}|\beta|^{2n}}{\pi n!}.
\label{em:Q_Fock}
\end{equation}
This distribution is positive but non-Gaussian. In the strong-field
semiclassical sampling, after neglecting the residual factor
\(e^{|\gamma|^2/2}\), the corresponding \(Q_n\)-weighted two-time kernel is
\begin{equation}
\nu_n(t,t')
=
2(n+1)\frac{\mathcal{E}_V^2}{\omega^2}
\cos\left[\omega(t-t')\right].
\label{em:nu_Fock}
\end{equation}
This kernel is the antinormally ordered second moment generated by the
\(Q\)-representation. Like the thermal kernel, it is stationary and therefore
produces no phase-sensitive \(2\omega\) variance modulation. Unlike the
thermal state, however, the Fock-state \(Q_n\) weight is non-Gaussian, so its
full fluctuation statistics is not determined by this two-time kernel alone;
higher moments are required to distinguish it from thermal noise, despite the
fact that both give only stationary second-order fluctuations.

\textit{Wave-packet reduction and TDSE implementation.---}
The monochromatic formulas above expose the phase-space structure. In the
calculations, the narrowband multimode IR pulse is projected onto an effective
wave-packet mode with envelope \(f(t)\), so the finite-pulse kernels are
obtained by multiplying the monochromatic kernels by \(f(t)f(t')\), as in
Eq.~\eqref{eq:kernel}. The stochastic process is then confined to the two
quadratures \(f(t)\cos(\omega t)\) and \(f(t)\sin(\omega t)\), and sampling an
entire noise history reduces to sampling a single complex amplitude \(\beta\).
Each realization propagates the electron in
\begin{equation}
A(t;\beta)
=
\frac{\mathcal{E}_V}{\omega}
f(t)
\left(
\beta e^{-i\omega t}
+
\beta^*e^{i\omega t}
\right),
\label{em:A_beta}
\end{equation}
with transition probability
\begin{equation}
P
=
\int\!\mathrm{d}^2\beta\,
\mathcal W_{\mathrm{eff}}(\beta)
\left|
\langle\psi_{eB}|
\hat U_{\beta}(t_f,t_i)
|\psi_{eA}\rangle
\right|^2,
\label{em:P_TDSE}
\end{equation}
where \(\mathcal W_{\mathrm{eff}}\) is the appropriate positive measure:
the Wigner function when it is positive, or the strong-field Husimi-\(Q\)
weight otherwise. The propagator is generated by
\begin{equation}
\hat H_{\beta}(t)
=
\frac{1}{2}
\left[
\hat{\bm p}
+
\bm\varepsilon A(t;\beta)
\right]^2
+
V_{\mathrm C}(\hat{\bm x}).
\label{em:H_beta}
\end{equation}
Thus the TDSE propagation remains a standard semiclassical streaking
calculation; the quantum-optical content enters through the phase-space weight
assigned to the classical IR amplitude.

This construction provides a general strong-field quantum-optical route to
semiclassical photon-traced electron dynamics. The influence functional first
selects the Wigner representation, which gives the exact fluctuation weight
whenever it is positive. Only when the Wigner function cannot serve as a
probability density do we switch to the always-positive Husimi-\(Q\)
representation, at the price of neglecting a vacuum-fluctuation-scale factor in
the sampling formula.
\end{widetext}
\end{document}